\documentclass[11pt]{article}

\usepackage{graphicx,fancyhdr,amsmath,amssymb,amsthm,subfig,url,hyperref}
\usepackage[margin=1in]{geometry}
\usepackage{enumitem}
\usepackage{apacite}
\usepackage{natbib}
\usepackage{amsmath,amssymb}
\usepackage[title]{appendix}
\usepackage{bbm}
\usepackage{mathtools}

\usepackage{mathtools}

\DeclarePairedDelimiter\floor{\lfloor}{\rfloor}

\usepackage{titlesec}
\titleformat{\chapter}[display]
  {\headingfont\centering}{\thechapter}{0pt}{\Huge\textbf}

\usepackage{tikz}
\usepackage{pgfplots}
 
\usepackage{xurl}

\usepackage{setspace}
\usepackage{textcomp}

\usepackage[graphicx]{realboxes}
\usepackage{graphicx} 
\graphicspath{{D:/My Documents/6015/essay2}}

\usepackage[affil-it]{authblk}

\author{Chan Tung Yu Marco}
\affil{University of Hong Kong}

\graphicspath{{figures/}}

\title{Expanding on Repeated Consumer Search Using Multi-Armed Bandits and Secretaries\footnote{HKU Fall Semester 2020; ECON 6077: Topics in Economics Research I - Industrial Organization, final project.}}

\begin{document}
\maketitle

\begin{abstract}
    We seek to take a different approach in deriving the optimal search policy for the repeated consumer search model found in \cite{fishman1995durability} with the main motivation of dropping the assumption of prior knowledge of the price distribution $F(p)$ in each period. We will do this by incorporating the famous multi-armed bandit problem (MAB). We start by modifying the MAB framework to fit the setting of the repeated consumer search model and formulate the objective as a dynamic optimization problem. Then, given any sequence of exploration we assign a value to each store in that sequence using Bellman equations. We then proceed to break down the problem into individual optimal stopping problems for each period which incidentally coincides with the framework of the famous secretary problem where we proceed to derive the optimal stopping policy. We will see that implementing the optimal stopping policy in each period solves the original dynamic optimization by `forward induction' reasoning.

\end{abstract}

\section{Introduction}

In the repeated consumer search model \citep{fishman1995durability} consumers are faced with a market selling the same product but with price dispersion. Consumers then have the objective of maximizing their own surplus by purchasing a unit of the product but at the lowest possible price, but are faced with the additional challenge of not knowing the prices in each store. They can resolve this lack of information by searching (exploration) which incurs a search cost $c$. Hence the main essence of the consumer's problem is to decide whether or not exploration is worthwhile. The setting is also a dynamic one, meaning that the consumer will have to repeat this purchasing process in every time period until the terminal period. The dynamic setting is used to portray the potential for long-term business relationships between firms and consumers. 

The model in \cite{fishman1995durability} proceeds its analysis with the assumption that consumers know the price distribution $F(p)$ beforehand in each period. We drop this assumption for the goal of deriving a search policy that captures more realism as this assumption is very hard to translate into reality. (How does one go about knowing the price distribution in each period? Why is this? What does this assumption represent?) Furthermore, the optimal search rule in \cite{fishman1995durability} seems very abstract in that it only describes the behavior of consumers given different transition probabilities rather than giving more concrete instructions to the consumer.  

To derive the optimal search policy without the prior knowledge assumption, we will incorporate the famous multi-armed bandit problem (MAB) which has found many applications even outside economics. In \cite{mccall1987sequential} they also use the MAB framework for designing a model for migration and sequential job search models, in which we aim to do something similar.

The MAB problem is used as it inherently assumes that agents have no information (though priors are allowed) about reward distributions.\footnote{We will give an overview of the MAB problem in later sections.} By formulating the search model into the MAB framework, we hope to be able to adopt and modify computational methods and results to provide us with further insight into a consumer's optimal search rule under relaxed assumptions and under a more general setting. It should also be noted that this paper's analysis is focused on the consumer, demand-side. Further analysis on the producer side will be required in order to gain any insight or implications this general setting may have on the market equilibrium.

\section{Literature Review}
This section will give a brief overview and summary of relevant literature and existing models.

\subsection{The Repeated Purchase Search Model \citep{fishman1995durability}}
Here we will introduce our starting point, the repeated purchase search model for a single good more formally, which is a natural extension of the price dispersion model in \cite{reinganum1979simple}.\footnote{\cite{reinganum1979simple} models the equilibrium of firms subject to productivity shocks (represented by marginal cost being randomly distributed) and consumers with imperfect information.} 

The model's set-up starts with a large number $n$ of identical consumers with individual demand for the single good $x = D(p)$. Each consumer will purchase a unit of the good every period $t=1,2,\ldots$ Then a consumer's surplus from purchasing at price $\hat{p}$ at time $t$ is, 

\begin{equation*}
S(\hat{p}) = \int_{\hat{p}}^\infty D(p) ~ \mathrm{d}p.
\end{equation*}

Consumers are assumed to only know the distribution of prices charged by each firm in the market but not the individual prices charged by each firm. Consumers are also allowed in each period to sequentially search and sample prices from firms they have not bought from in the previous period at a cost $c>0$.\footnote{In contrast, a consumer can costlessly learn the current price of any seller she bought from in the last period. This can be interpreted as the consumer starting every new time period at the store they bought from in the last period. The idea is that consumers have all information about sellers they previously bought from but retains no information about firms they merely visited, or bought from more than a period ago.}

The net period surplus of a consumer who has searched $n_t$ stores and buys at price $p_t$ is given by $S(p_t) - n_t c$, giving us the sum of discounted surplus,

\begin{equation}
V(t) = \sum_{\tau = t}^\infty \delta^{\tau-t} \Big[S(p_\tau)-n_\tau c \Big]
\label{discounted net surplus}
\end{equation}

with discount factor $\delta \in (0,1)$.

We simplify the cost structure for firms so that marginal cost $\omega$ can take two distinct values $\{\omega_L, \omega_H\}$ where $\omega_L \leq \omega_H.$ We also assume that marginal cost of firms satisfy the Markov property, that is,
\begin{align*}
\beta_H &= \mathrm{P}(\omega_t = \omega_H \mid \omega_{t-1} = \omega_H)  \\
\beta_L &= \mathrm{P}(\omega_t = \omega_L \mid \omega_{t-1} = \omega_L). 
\end{align*}

For simplification $\beta = \beta_L = \beta_H$ is assumed, so $\beta$ is the persistence probability of a firm to maintain its current productivity level. That is, in every given period, a firm's state of being a high or low cost is subject to change as dictated by $\beta$. This cost state is determined at the start of every period will remain unchanged throughout the entire period.

Then assuming the system has a steady state, it can be shown that the number of low cost and high cost firms will be equal regardless of the value of $\beta$. 

Then given state of a firm, it can be shown that in equilibrium low cost firms will charge $p_L = p_L^m$ and high cost firms will charge $p_H = \min\{p_r,p_H^m\}$, where $p^m$ is the profit maximizing (monopoly) price\footnote{Maximizing expected discounted profits $\Pi_t = \mathrm{E}\Big[\sum_{\tau = t}^{\infty} \delta^{\tau - t} D(p_\tau)(p_\tau-\omega_\tau)\Big]$} $p_r$ is the (unknown) reservation price of consumers. It should be noted that the reservation price $p_r$ is what dictates the consumers' search rule i.e. if $p_j > p_r$ then continue searching, otherwise settle and purchase.

In a steady state equilibrium  we assume that consumers will engage in optimal sequential search to maximizes their discounted net surplus (\ref{discounted net surplus}). Then a steady state equilibrium will be characterized by a (joint) price distribution\footnote{Note that $F$ here can be formulated as a binomial distribution, given that $p_j$ are all independent and can be formulated as Bernoulli random variables.} $F(p_1,p_2,\ldots,p_M)$ given $M$ firms in the market that is unchanging over time and satisfies,

\begin{enumerate}
\itemsep 0.1em 
\item Given $F(p)$, the consumer's search rule maximizes their discounted net surplus (\ref{discounted net surplus})
\item Given $F(p)$, $\omega_j$, number of customers last period and consumers' search rules, each firm maximizes their discounted expected profits. 
\item The firms' profit maximization gives rise to the same $F(p)$ given in 1. and 2.
\end{enumerate}

As in \cite{reinganum1979simple}, it can be shown that $p_L < p_r$. That is, it is not worth for consumers to search until the lowest price $p_L$ is found if they have found a price sufficiently close to $p_L$. 

First we assume that at the steady state equilibrium, high cost firms are constrained and cannot charge the monopoly price, i.e. $p_H = p_r$. Then writing out the expected surplus $V_L$ for a consumer receiving low price $p_L$ at time $t$ together with the fact that at steady state, there is an equal number of high and low cost firms, we can formulate the system of equations

\begin{equation}
\begin{cases}
V_H &= -c + \frac{1}{2}V_H + V_L = S(p_r) + \delta \big[\beta V_H + (1-\beta)V_L\big] \\
V_L &= S(p_L) + \delta \big[\beta V_L + (1-\beta)V_H\big]
\end{cases}
\end{equation}

where $V_H$ is the expected discounted surplus from search continuation\footnote{$p_H = p_r$ implies that consumers are indifferent between settling for $p_H$ and continuing their search.}.

Solving the above system we will arrive at,
\begin{equation}
S(p_r) = S(p_L) + (4\beta\delta -2\delta - 2)c.
\label{reservation surplus}
\end{equation}

To understand this result, consider the two extreme cases, when $\beta = 1$ (or equivalently $\beta = 0$) and $\beta = \frac{1}{2}.$ Since all firms are `fixed' (low price firms will indefinitely offer low prices), given that $p_r$ is the current price on hand, the lifetime benefit of finding a low price firm is

\begin{equation}
\sum_{\tau = t}^\infty \big[S(p_L)-S(p_r)\big] = \frac{S(p_L)-S(p_r)}{1-\delta}.
\label{lifetime benefit of low price}
\end{equation}

Also notice that when we substitute $\beta = 1$ into (\ref{reservation surplus}) we get 

\begin{equation}
c = \frac{1}{2}\cdot\frac{S(p_L)-S(p_r))}{1-\delta}.
\label{indifference condition beta 1}
\end{equation}

Recall that since there is an equal number of firms, the expected benefit of searching is $\frac{1}{2}$ multiplied by (\ref{lifetime benefit of low price}). Hence, the consumer will set reservation price $p_r$ so as to equate the expected benefit of searching with search cost $c$, given by (\ref{indifference condition beta 1}). The intuition being that search is most valuable when $\beta=1$. Since all firms are fixed, the consumer can immediately pin down their discounted lifetime benefit of all future expenditures when deciding between settling with their current price and searching.  

On the other hand, when $\beta = 0$, there is no persistence of the firm's state across periods. The intuition here is since firms now have equal probabilities of changing state, together with the fact that the number of low and high cost firms are half-half at steady state, we can see that searching has zero value as it only incurs a search cost $c$. Then for a consumer who has $p_r$ on hand, if he were to find a low cost firm his benefit of search would be $\big[S(p_L)-S(p_r)\big]$ which would occur with probability $\frac{1}{2}$ given equal number of high and low cost firms at steady state. Similarly, when we substitute $\beta = 0$ into (\ref{reservation surplus}) we get 

\begin{equation}
c = \frac{1}{2}\big[S(p_L)-S(p_r)\big]
\label{indifference condition beta 1/2}
\end{equation}

where consumers will similarly choose $p_r$ to satisfy the indifference condition (\ref{indifference condition beta 1/2}).

We can see that in the end, although cost distribution is invariant to persistence probability $\beta$, the price distribution is.

\subsection[Overview of The Multi-Armed Bandit (MAB) Problem]{Overview of The Multi-Armed Bandit (MAB) Problem\footnote{\cite{weng2018bandit}, \cite{slivkins2019introduction}.}}
The multi-armed bandits (MAB) problem is a classic reinforcement learning problem introduced by \cite{thompson1933likelihood}. The problem exemplifies the trade-off dilemma between exploration and exploitation (very much like our search theory models.) An armed-bandit is a nickname given to casino slot machines where one can pull the arm to receive a random reward (essentially a lottery.)  

The most basic framework of the MAB problem are the `stochastic bandits'. The problem is formulated as such. There are $K$ armed-bandits, each of which provides a random reward $r^j \in [0,1]$ distributed by $F^j$ which are i.i.d.\footnote{Independently, identically distributed} for all $j=1,2,\ldots,K$. Thus, the $K$ armed-bandits can be summarized by the vector of mean rewards $(\mu^1, \ldots, \mu^K).$ 

Define $\mathcal{A}_t$ as the set of actions\footnote{Where in our case we have $\mathcal{A}_1$ = $\mathcal{A}_2$ = \ldots = $\mathcal{A}_T$.} at time $t$ which describes the interaction with a single bandit at time $t$. So, if action $a_t$ is taken at time $t$ on bandit $j$, then the `action value' of $a_t$ is the mean reward $Q(a_t)=\mu^j$ and her realized reward will be $\pi(a_t)=r^j$, where $a_t$ is the action taken\footnote{$a_t$ can be interpreted as the arm chosen to be pulled at time $t$.} and $\pi$ is the reward function. The agent's objective is to maximize his total reward over $T$ rounds. Of course, in this model the agent is assumed to not know the reward distributions $F^1,F^2,\ldots,F^K$. 

Thus the dilemma arises, where in every round the agent must choose between finding a different arm to pull (exploration/search) and continuing to pull from the same bandit. 

The standard measure of performance of a strategy/algorithm in the MAB problem or equivalently, an alternate formulation of the objective in the MAB problem is to minimize what is called `regret'. Regret is measured using the \textit{best-arm} as a benchmark to compare by. Let $\mu^* = \max_j \mu^j$, that is $\mu^*$ is the highest mean reward (or the mean reward of the optimal arm). 

Then define regret $R$ at time $t$ as,

\begin{equation}
R(t) = t\mu^* - \sum_{\tau = 1}^t Q(a_\tau)
\label{regret}
\end{equation}

where $Q(a) = \mathrm{E}\big[\pi(a)\big]$ is the action value of $a$.

Generally, there are three ways the agent can go about forming his strategy.
\begin{enumerate}
\itemsep 0.1em 
\item No exploration (trivial case)
\item Random exploration
\item Strategic exploration with preferences over uncertainty
\end{enumerate}

There are several famous algorithms ($\varepsilon$-greedy algorithm, Upper Confidence Bounds (UCB)) the agent can choose to implement which fall into one of the above three categories. Also, the problem has been approached using dynamic programming methods\footnote{See \cite{mccall1987sequential} and \cite{gittins1974dynamic}.} which will be the main approach used in this paper. 

We can already see the many parallels between the MAB problem and the consumer search model, with the main difference being that the agent has no information regarding the reward distributions of each bandit. Another striking difference is that the repeated purchase model \citep{fishman1995durability} as formulated in the MAB framework can be seen as a more 'forgiving' version of the classical MAB, in that consumers can sample (at a search cost $c$) whereas in the the MAB, consumer would choose a seller and immediately buy at the offered price.

\section[Repeated Purchase and the MAB]{Repeated Purchase and the MAB}
First, we will formulate the repeated search model \citep{fishman1995durability} using Bernoulli bandits.\footnote{Bandits where the state space is $\{0,1\}$.} The consumers can be represented by a single agent, and the $M$ firms by $M$ Bernoulli bandits (or $M$-armed Bernoulli bandits.) 

Now for the store chosen at time $t$, $j(t)$ define the state of store $j(t)$,

\begin{equation}
x\big[j(t)\big] =
    \begin{cases}
    1 ~\ ~\ \text{if store is low cost at $t$} \\
    0 ~\ ~\ \text{if store is high cost at $t$}.
    \end{cases}
    \label{state of store}
\end{equation} 

Recall that the state $x\big[j(t)\big]$ follows a (stationary) Markov process and we found that under a steady-state we have that the probability of finding a high/low cost store is $\frac{1}{2}$. 

Then the reward function will be defined as 
\begin{equation}
\pi\Big(x\big[j(t)\big]\Big) =
\begin{cases}
S(p_L) ~\ \text{if $x\big[j(t)\big] = 1$} \\
S(p_H) ~\ \text{if $x\big[j(t)\big] = 0$}
\end{cases}
\label{reward function for single good bernoulli bandit}
\end{equation}

That is, the reward function depends on the store chosen and its state at time $t$. In terms of the bandit framework, $j(t)$ is the arm chosen at time $t$ and the reward realized is $r_j = \pi\Big(x\big[j(t)\big]\Big)$ which we will sometimes shorten to $\pi\big[x(t)\big]$. 

One other important item to mention is that given the reward function formulation in (\ref{reward function for single good bernoulli bandit}), we are assuming that on the producer/firm side, they are behaving optimally and that prices reflect their productivity state as was in \cite{fishman1995durability}, in that low cost firms offer $p_L$ and high cost firms offer $p_H$ with $p_L < p_H$.

Then after incorporating search cost $c$ the consumer's objective is the dynamic optimization problem,

\begin{equation}
    \max_{\{n(t),j(t)\}_{t=0}^\infty} \sum_{t=0}^\infty \delta^t\Bigg[\pi\Big(x\big[j(t)\big]\Big)-c\cdot n(t)\Bigg]
    \label{single good bernoulli bandit objective}
\end{equation}

with discount rate $\delta \in (0,1)$. In words, this means that in every period the consumer will want to optimally choose the best store (arm) $j(t)$ given the optimal number of stores searched $n(t)$ in each period $t$.

This formulation of the search problem as a MAB problem follows a very similar form to that in \cite{mccall1987sequential} where they incorporate dynamic programming methods together with Gittins' index introduced in \cite{gittins1974dynamic}, the decision rule being `always play the bandit with the largest index' in which we take a similar but different approach.\footnote{Although we will not be assigning each store with a Gittins' index, we will be assigning values on each store as shown in the next section. The policy we derive is also an optimal stopping rule.}

\begin{figure}[h]
    \centering
\begin{tikzpicture}[>=stealth]


\begin{axis}[
axis x line =center,
axis y line=center,
xtick={0,1,2,3,4,5},
ytick={0,1,2,3,4,5},
xlabel={Store $j$},
ylabel={Time $t$},
xlabel style={below right},
ylabel style={above left},
xmin = 0,
xmax = 6,
ymin= 0,
ymax = 6]
\end{axis}

	\filldraw [black] (1.15,0) circle (3pt);
	\filldraw [orange] (2*1.15,0) circle (3pt);
	\filldraw [black] (3*1.15,0) circle (3pt);
	\filldraw [black] (4*1.15,0) circle (3pt);
	\filldraw [black] (5*1.15,0) circle (3pt);

	\filldraw [black] (1.15,0.9) circle (3pt);
	\filldraw [orange] (2*1.15,0.9) circle (3pt);
	\filldraw [orange] (3*1.15,0.9) circle (3pt);
	\filldraw [black] (4*1.15,0.9) circle (3pt);
	\filldraw [orange] (5*1.15,0.9) circle (3pt);
	
   	\filldraw [black] (1.15,1.9) circle (3pt);
   	\filldraw [black] (2*1.15,1.9) circle (3pt);
   	\filldraw [orange] (3*1.15,1.9) circle (3pt);
   	\filldraw [black] (4*1.15,1.9) circle (3pt);
   	\filldraw [black] (5*1.15,1.9) circle (3pt);

	\filldraw [black] (1.15,2.85) circle (3pt);
	\filldraw [black] (2*1.15,2.85) circle (3pt);
	\filldraw [black] (3*1.15,2.85) circle (3pt);
	\filldraw [black] (4*1.15,2.85) circle (3pt);
	\filldraw [black] (5*1.15,2.85) circle (3pt);

	\filldraw [black] (1.15,3.8) circle (3pt);
	\filldraw [black] (2*1.15,3.8) circle (3pt);
	\filldraw [black] (3*1.15,3.8) circle (3pt);
	\filldraw [black] (4*1.15,3.8) circle (3pt);
	\filldraw [black] (5*1.15,3.8) circle (3pt);
	
	\filldraw [black] (1.15,4.75) circle (3pt);
	\filldraw [black] (2*1.15,4.75) circle (3pt);
	\filldraw [black] (3*1.15,4.75) circle (3pt);
	\filldraw [black] (4*1.15,4.75) circle (3pt);
	\filldraw [black] (5*1.15,4.75) circle (3pt);

\draw[->] [blue] (1*1.15,0) -- (1*1.15,0.7) node[left, yshift=-0.3cm] {};
\draw[->] [blue] (2*1.15,0) -- (2*1.15,0.7) node[left, yshift=-0.3cm] {};
\draw[->] [blue] (3*1.15,0) -- (3*1.15,0.7) node[left, yshift=-0.3cm] {};
\draw[->] [blue] (4*1.15,0) -- (4*1.15,0.7) node[left, yshift=-0.3cm] {};
\draw[->] [blue] (5*1.15,0) -- (5*1.15,0.7) node[left, yshift=-0.3cm] {};

\draw[->] [blue] (1*1.15,0.9) -- (1*1.15,1.7) node[left, yshift=-0.3cm] {};
\draw[->] [blue] (2*1.15,0.9) -- (2*1.15,1.7) node[left, yshift=-0.3cm] {};
\draw[->] [blue] (3*1.15,0.9) -- (3*1.15,1.7) node[left, yshift=-0.3cm] {};
\draw[->] [blue] (4*1.15,0.9) -- (4*1.15,1.7) node[left, yshift=-0.3cm] {};
\draw[->] [blue] (5*1.15,0.9) -- (5*1.15,1.7) node[left, yshift=-0.3cm] {};

\draw[->] [blue] (1*1.15,1.9) -- (1*1.15,2.7) node[left, yshift=-0.3cm] {};
\draw[->] [blue] (2*1.15,1.9) -- (2*1.15,2.7) node[left, yshift=-0.3cm] {};
\draw[->] [blue] (3*1.15,1.9) -- (3*1.15,2.7) node[left, yshift=-0.3cm] {$P_{x,y}$};
\draw[->] [blue] (4*1.15,1.9) -- (4*1.15,2.7) node[left, yshift=-0.3cm] {};
\draw[->] [blue] (5*1.15,1.9) -- (5*1.15,2.7) node[left, yshift=-0.3cm] {};

\draw[->] [blue] (1*1.15,2.9) -- (1*1.15,3.6) node[left, yshift=-0.3cm] {};
\draw[->] [blue] (2*1.15,2.9) -- (2*1.15,3.6) node[left, yshift=-0.3cm] {};
\draw[->] [blue] (3*1.15,2.9) -- (3*1.15,3.6) node[left, yshift=-0.3cm] {};
\draw[->] [blue] (4*1.15,2.9) -- (4*1.15,3.6) node[left, yshift=-0.3cm] {};
\draw[->] [blue] (5*1.15,2.9) -- (5*1.15,3.6) node[left, yshift=-0.3cm] {};

\draw[->] [blue] (1*1.15,3.9) -- (1*1.15,4.6) node[left, yshift=-0.3cm] {};
\draw[->] [blue] (2*1.15,3.9) -- (2*1.15,4.6) node[left, yshift=-0.3cm] {};
\draw[->] [blue] (3*1.15,3.9) -- (3*1.15,4.6) node[left, yshift=-0.3cm] {};
\draw[->] [blue] (4*1.15,3.9) -- (4*1.15,4.6) node[left, yshift=-0.3cm] {};
\draw[->] [blue] (5*1.15,3.9) -- (5*1.15,4.6) node[left, yshift=-0.3cm] {};

\draw[->] [red] (3*1.15,1.9) -- (4.4,1.9) node[below, xshift=-0.4cm] {$c$};

\end{tikzpicture}
    \caption{State space diagram}
    \label{fig:phase space}
\end{figure}
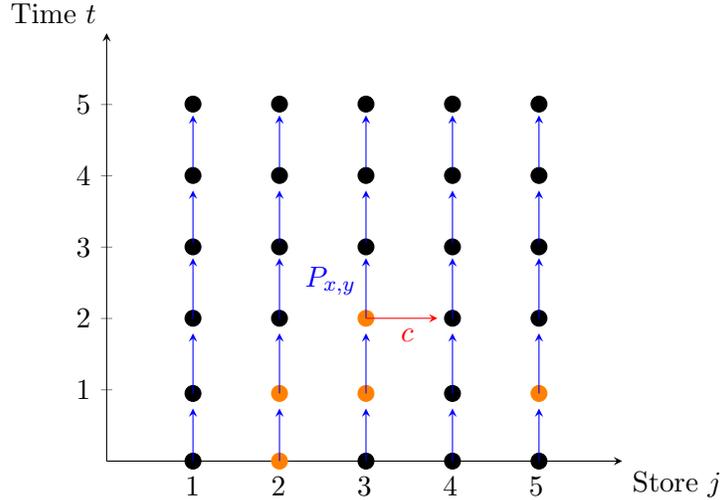

One major difference between our context and other MAB formulations is that untouched bandits do not `freeze'. That is, marginal cost of the store are still subject to change (via. Markov process) regardless of whether or not that store has been visited. Figure \ref{fig:phase space} depicts a visualization of the problem, where each black circle represents store $j$'s (unobserved) state at time $t$, an orange circle represents a store who's state is observed at $t$, $P_{x,y}$ is the transition probability and $c$ is the search cost. The blue arrows represent the Markov transitions of each store going forward in time whilst the red arrow indicates search across stores in each period.\footnote{A reminder that we are not considering a spatial model, so a consumer can travel to any single store starting from any store at the search cost $c$.} 

\subsection{Dynamic Programming Framework}
To re-iterate the setup, there are $M$ stores in the market, each of which are subject to changes in productivity via a Markov process represented by the transition matrix $P$ given our two-state setup,

\begin{equation}
    P = \begin{pmatrix} \beta_L & 1-\beta_L \\ 1-\beta_H & \beta_H \end{pmatrix}.
    \label{transition matrix}
\end{equation}

One important thing to note is that although the consumers are unaware of the price distribution $F(p)$ in each time period, we will maintain the assumption that they are aware of the transition probabilities, i.e. the transition matrix $P$.\footnote{Note that consumers' knowledge of the transition probabilities may lead them to deriving the stationary distribution of prices, if there is one to be found. This is however to do with an individual store and the probability distribution of the state that particular store given a certain time $t$.} This can be interpreted as consumers being aware of external market factors influencing productivity shocks of stores. 

Similar to \cite{gittins1974dynamic} and \cite{mccall1987sequential}, to formulate the problem as a dynamic program the consumer will assign a value to each store. To see how, consider the following illustration.

First, at the initial period $t=0$, let the consumer's initial position be at one of the $M$ stores; that is she gets to explore the first store free of charge.\footnote{This assumption smoothens the framework when formulating the Bellman equations.} Then before the start of each period, the consumer predetermines the order in which she explores any additional stores.\footnote{Any store explored after the initial store is considered `additional'.} Since the price distribution is unknown to her, the sequential order she chooses for exploration is inconsequential. Now denote the order of the stores in which she plans to explore as $\{1,2,\ldots,M-1\}$. Then starting at $t=0$, we define the value placed on each store $j=0,1,2,\ldots,M-1$ as,

\begin{equation}
    \begin{split}
        V_0 \Big[x_1(0)\Big] &= \pi \Big[x_1(0)\Big] + \delta \Bigg[\sum_{s=1}^2 P_{x_1(0), s} \cdot V_0(s)\Bigg] \\
        V_1 \Big[x_1(0), x_2(0)\Big] &= \max \Bigg\{\pi\Big[x_1(0)\Big],\pi \Big[x_2(0)\Big]\Bigg\} - c + \delta \Bigg[\sum_{s=1}^2 P_{x_2^*(0), s} \cdot V_0(s)\Bigg] \\
        &\vdots \\
        V_{M-1} \Big[x_1(0),\ldots, x_{M}(0)\Big] &= \max \Bigg\{\pi\Big[x_1(0)\Big],\ldots,\pi \Big[x_{M}(0)\Big]\Bigg\} - (M-1)c + \delta \Bigg[\sum_{s=1}^2 P_{x_{M}^*(0), s} \cdot V_0(s)\Bigg]
    \end{split}
    \label{dynamic store values}
\end{equation}

where the $j$ subscript on $V_j$ denotes the order in which that additional store was visited, i.e. the $j^{\text{th}}$ additional store visited in the order of exploration $\{1,2,\ldots,M-1\}$, $x_m(0)$ denotes the state of store $m$ and $x_m^*(0) = \mathrm{argmax}_{k \in \{1,2,\ldots,m\}} \Big\{ \pi\big[x_k(0)\big] \Big\}$ at $t=0$.\footnote{A short reminder that we assume the consumer is able to return to any store she has sampled at no cost but only for that period.} 

There are several important things to note regarding the formulation in (\ref{dynamic store values}). First is that this formulation is valid for any period $t$. Next is that the initial dynamic optimization problem in (\ref{single good bernoulli bandit objective}) required choices on both number of stores to explore $n(t)$ and the store to purchase from $j(t)$ in each period $t$. By assigning values to each store as done in (\ref{dynamic store values}) we have broken down the problem into a sequence of individual optimal stopping problems. 

To see how, first notice that the formulation of the values in (\ref{dynamic store values}) takes in the stores observed thus far and provides the maximum expected value for the current period and the subsequent, taking search costs into account.\footnote{Observe that the $V_0(s)$ term inside the expectation portion depicts that the value is derived from zero additional searches as the consumer is assumed to appear in the store she purchased from last period.} Then notice that at $t=0$ (or any subsequent time period) the consumer is first faced with the choice of whether to settle for value $V_0 = V_0\big[x_1(0)\big]$ or go onto the first store in the predetermined order of search $\{1,2,\ldots,M-1\}$ to realize value $V_1$. Then by construction, if she accepts the value $V_1$, there is no reverting back to the value $V_0$ as the search cost $c$ is built into the value $V_1$. Similarly, she can choose whether to discard $V_1$ to realize $V_2$ and so on until $V_{M-1}$. We now start to see that this formulation is precisely that of the famous secretary problem (also known as the marriage problem) that was introduced in the early 1960's by Martin Gardner.\footnote{See \cite{ferguson1989solved} for a history and review of the problem.} 

\subsection[The Optimal Stopping (Secretary) Problem]{The Optimal Stopping (Secretary) Problem\footnote{We use the dynamic programming approach found in \cite{beckmann1990dynamic} to derive the optimal stopping policy.}}
We will first formulate the initial optimal stopping problem at $t=0$ and derive its optimal policy.\footnote{This can be applied to any arbitrary period $t$.} At the start of $t=0$ the consumer is faced with the sequence of values $\{V_0, V_1, V_2, \ldots, V_{M-1}\}$ where only $V_0$ is known initially. Had the consumer been clairvoyant, she would be able to rank the stores according to their values from greatest to least. Alas, her objective is to; given a predetermined sequence of exploration $\{1,2,\ldots,M-1\}$; find an optimal stopping policy that maximizes her received value. The optimal stopping policy will take the following form: `Pass through a \textit{certain number} of stores, and after that pick the first store that yields the highest value so far.'

Before we proceed with the derivation, we will introduce some notation make some simplifications for convenience. For the purpose of deriving the optimal stopping policy, we focus our attention only to the rank order of the stores' values $V_j$, not to the actual values themselves. So, the consumer maximizing her probability of finding the best store is equivalent to maximizing her expected value received.\footnote{Suppose after receiving a value $V_j$ the consumer is convinced that there is a high chance that $V_j$ is the highest value. So we can see that maximizing the probability of finding the highest value is indeed the same as maximizing her received value.} Hence, for this section the term `value' will also refer to the expected probability of finding the best store.  

\begin{itemize}
    \item Let $m$ be the number of stores visited/sampled. So the consumer starts with $m=1$ which is the store with value $V_0$.\footnote{If $j$ is the index of search order, then $m = j+1$. So after exploring $M-1^{\text{th}}$ store, $m=M$.} 
    \item Let $n$ be the number of stores not yet visited, so $M = m+n$. 
    \item We say that a store is `\textit{viable}' if the store's value is the highest seen so far. 
    \item Define $Y_m$ to be the value when $m$ stores have been visited and $V_m$ has been discarded. 
    \item Define $U_m$ to be the value when $m$ stores have been visited and the $m^{\text{th}}$ store is \textit{viable}.
\end{itemize}

Consider the case when the $m^{\text{th}}$ is not chosen, then the value for the consumer is,

\begin{equation}
    \begin{split}
        Y_m &= \mathrm{P}\Big\{m+1^{\text{th}} ~\ \text{store is not viable}\Big\}Y_{m+1} + \mathrm{P}\Big\{m+1^{\text{th}} ~\ \text{store is viable}\Big\}U_{m+1} \\
        &= \Big(\frac{m}{m+1}\Big)Y_{m+1} + \Big(\frac{1}{m+1}\Big)U_{m+1}. 
    \end{split}
    \label{Ym}
\end{equation}
 
Now consider the case that the $m^{\text{th}}$ store is viable. There is a decision either to choose or discard $V_m$. Then if the $m^{\text{th}}$ store is chosen then the value is $\mathrm{P}\big\{m^{\text{th}} ~\ \text{is the best}\big\} = m/M$.\footnote{In this case, $m/M$ can be interpreted as the `search termination' value.} Since $Y_m$ has already been established we have,

\begin{equation}
    U_m = \max \Big\{\frac{m}{M}, Y_m \Big\}.
    \label{Um}
\end{equation}

Using backward induction, we initiate with the last store by setting $m= M$ so we have $Y_{M} = 0$ by definition.\footnote{Discarding the final value offered by the last store results in 0 probability of finding the best value.} Then we have $U_{M} = \max\{1,0\} = 1.$ 

Then by (\ref{Ym}), 
\begin{equation}
        Y_{M-1} = \Big(\frac{M-1}{M}\Big)Y_{M} + \Big(\frac{1}{M}\Big)U_M = \frac{1}{M}
\end{equation}

and,
\begin{equation}
    U_{M-1} = \max \Big\{\frac{M-1}{M}, \frac{1}{M} \Big\}  = \frac{M-1}{M}.
\end{equation}
 
Continuing to $m = M-2$, 
 
\begin{equation}
    \begin{split}
        Y_{M-2} &= \Big(\frac{M-2}{M-1}\Big) Y_{M-1} + \Big(\frac{1}{M-1}\Big)U_{M-1} \\[0.2cm]
        &= \frac{1}{M}\Big(\frac{M-2}{M-1} + 1 \Big) \\[0.2cm]
        &= \frac{1}{M}\Big(\frac{M-2}{M-1} + \frac{M-2}{M-2}\Big) =  \frac{M-2}{M}\Big(\frac{1}{M-1} + \frac{1}{M-2}\Big)
    \end{split}
\end{equation}

and,

\begin{equation}
    U_{M-2} = \max \bigg\{\frac{M-2}{M}, \frac{M-2}{M}\Big(\frac{1}{M-1} + \frac{1}{M-2}\Big) \bigg\}.
\end{equation}
~\\[0.1cm]
Starting to see a pattern emerge, given $m=M-n$ for $n=1,2,\ldots,M-1$ we have,
\begin{equation}
    \begin{split}
        Y_{m} &= \frac{m}{M}\Big(\frac{1}{m} + \frac{1}{m+1} + \ldots + \frac{1}{M-1}\Big) \\[0.2cm]
        U_{m} &= \frac{m}{M} \max\Big\{1, \frac{1}{m} + \frac{1}{m+1} + \ldots + \frac{1}{M-1}\Big\}. 
    \end{split}
    \label{Ym, Um general form}
\end{equation}
 
\subsection[Deriving the Optimal Stopping Policy]{Deriving the Optimal Stopping Policy\footnote{\cite{beckmann1990dynamic}}}

From (\ref{Ym, Um general form}) we can see that (a) $\sum_{k=m}^{M-1} \frac{1}{k}$ is decreasing in $m$, (b) $\sum_{k=m}^{M-1} \frac{1}{k} < 1$ implies that $U_m > Y_m$, that is; the value of the $m^{\text{th}}$ store being viable is greater than the value received by skipping the $m^{\text{th}}$ store.

Let $m = m^*$ be a critical cut-off point such that the optimal policy will choose the first viable store $m \geq m^*$. As such $m=m^*$ will solve,

\begin{equation}
    \sum_{k=m^*}^{M-1} \frac{1}{k} \leq 1 < \sum_{k=m^*-1}^{M-1} \frac{1}{k}.
    \label{m* condition}
\end{equation}

That is, the $m^*-1^{\text{th}}$ store is the last store such that,
\begin{equation}
    \frac{m^*-1}{M}\sum_{m^*-1}^{M-1}\frac{1}{k} > \frac{m^*-1}{M}.
    \label{value of rejecting m^*-1 > prob that m*-1 is best}
\end{equation}

The LHS of (\ref{value of rejecting m^*-1 > prob that m*-1 is best}) is the value of rejecting the $m^*-1^{\text{th}}$ store and the RHS is the probability that $m^*-1^{\text{th}}$ is the best store. So, $m=m^*$ will be the first store where the value of rejecting $m=m^*$ is greater or equal to the probability that $m=m^*$ is the best store. Then it follows that $Y_{m^*-1}$ is the value received by the optimal stopping policy.\footnote{See appendix \ref{Appendix A} to the derivation of (\ref{value of optimal policy}).} 

\begin{equation}
    Y_{m^*-1} = \frac{m^*-1}{M}\sum_{k=m^*-1}^{M-1}\frac{1}{k}.
    \label{value of optimal policy}
\end{equation}

It can also be shown that $U_0 = Y_0 = U_1 = Y_1 = \ldots = U_{m^*-1} = Y_{m^*-1}$ by noting the fact that $m^*-1$ is the last store such that $Y_{m^*-1} > \frac{m^*-1}{M}.$

We can approximate the solution $m^*$ by using,

\begin{equation}
    \int_{m^*}^M \frac{1}{x} ~ \mathrm{d}x \approx \sum_{k=m^*}^{M-1}\frac{1}{k} \leq 1
    \label{approximation}
\end{equation}

to give us the optimal policy's condition on critical cut-off $m^*$, 
\begin{equation}
    \ln\Big(\frac{M}{m^*}\Big) \approx 1 \Rightarrow \frac{m^*}{M} \approx \frac{1}{e}.
    \label{optimal policy condition}
\end{equation}

That is, the optimal stopping policy's probability of finding the best store is $1/e$.\footnote{The $1/e$ solution is a well-known result which was demonstrated in \cite{derman1970finite} and others (see \cite{ferguson1989solved} for details on the history.}   

Looking closer at (\ref{approximation}), we would notice that actually,

\begin{equation}
        \int_{m^*}^M \frac{1}{x} ~ \mathrm{d}x < \sum_{k=m^*}^{M-1}\frac{1}{k} \leq 1 
\end{equation}

implying, 
\begin{equation}
    m^* = \floor*{\frac{M}{e}} \Rightarrow ~ \frac{M}{m^*} < e   
\end{equation}

where $\floor*{\cdot}$ is the floor function.

Then from (\ref{approximation}) we see that $Y_m \approx \frac{m}{M} \ln\big(\frac{M}{m}\big)$. To get a better understanding of the result (\ref{optimal policy condition}) write $Y_m$ as a function of $m$ with parameter $M$,

\begin{equation}
    Y(m;M) = \frac{m}{M} \ln\Big(\frac{M}{m}\Big).
\end{equation}

\begin{figure}[h]
\begin{center}
\begin{tikzpicture}[scale=6]

  \draw[->] (0, 0) -- (1.1, 0) node[right] {$m$};
  \draw[->] (0, 0) -- (0, 1) node[above] {$Y(m;M)$};
  
  \draw[blue] plot [domain=0.001:1,samples=100] (\x,{\x/1*ln(1/\x)});
  \node at (0.8,0.3) {\footnotesize $Y(m;5)$};
  \node at (0.15,0.45) {\footnotesize $Y(m;2)$};
  
  \draw[red] plot [domain=0.001:0.3,samples=100] (\x,{\x/0.3*ln(0.3/\x)});
  
 \draw[dashed] (0,0.36787) -- (1,0.36787);
 
 \draw[dashed] (0.1104,0) -- (0.1104,0.36787);
 \draw[dashed] (0.3679,0) -- (0.3679,0.36787);

 \foreach \y in {0.36787} 
 \draw (1pt,\y cm) -- (-1pt,\y cm) node[anchor=east] {\footnotesize $\frac{1}{e}$};
 
\end{tikzpicture}
\end{center}
\caption{Visualization of $Y(m;M)$ across $M$ values}
\label{Y visualization}
\end{figure}
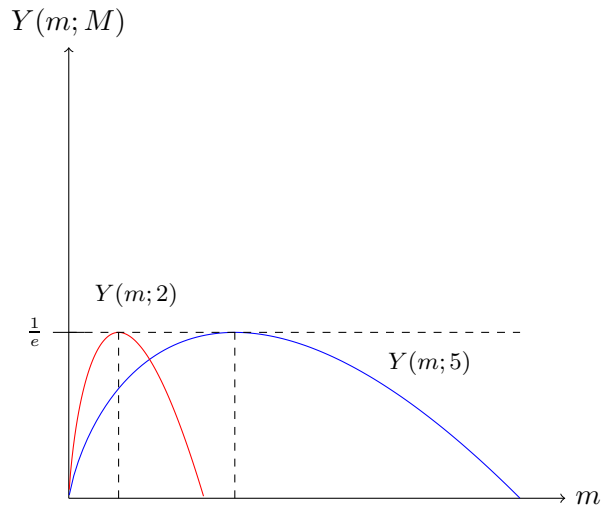

Then notice that the optimal $Y(m;M)$ value is decreasing in $m$ as $\frac{m}{M}$ increases (say via $M$ decreasing.) Figure \ref{Y visualization} shows a visualization of how the optimal $m$ value changes as $M$ increases. From figure \ref{Y visualization} we can see that as $\frac{m}{M}$ increases, the optimal cutoff $m^*$ decreases, hence the idea of accepting the first viable value from store $m^*$ onward.

Table \ref{table 1}  shows the optimal stopping policy's $m^*$ cut-off values and $Y_0$ values for varying number of stores $M$. We can see that the value $Y_0$ tends to and fluctuates around $1/e$ for a high enough number of stores $M$. 

Finally we may conjecture that implementing this optimal stopping policy every period is indeed an optimal strategy. Looking back at the value formulation in (\ref{dynamic store values}), we can choose to apply it to any arbitrary time period $t$. By doing so, if the consumer were to find the optimal value in that period denoted $V^*(t)$, that would mean that the current surplus gained from the store plus the expected value next period less the search cost is the highest amongst all stores at $t$. Using `forward induction' reasoning we can see that by virtue of the dynamic program formulation, the desirability of an action in the present is influenced by what may happen in the future. As a result, the optimal policy that maximizes the probability of finding the store with the maximum value $V^*(t)$ for every period $t$ is indeed optimal.

\clearpage
 
 \begin{table}[h] \centering 
   \caption[Optimal Stopping Policy Values]{Optimal Stopping Policy Values\protect\footnotemark} 
  \label{table 1} 
\begin{tabular}{@{\extracolsep{5pt}} ccccc} 
\\[-1.8ex]\hline 
\hline \\[-1.8ex] 
$M$ & $M/e$ & $m^*$ & $m^*-1$ & $Y_{0}$\\ 
\hline \\[-1.8ex] 
$1$ & $0.368$ & $0$ & $0$ & $1$ \\ 
$2$ & $0.736$ & $0$ & $0$ & $0.500$ \\ 
$3$ & $1.104$ & $1$ & $0$ & $0.500$ \\ 
$4$ & $1.472$ & $1$ & $0$ & $0.458$ \\ 
$5$ & $1.839$ & $1$ & $0$ & $0.417$ \\ 
$6$ & $2.207$ & $2$ & $1$ & $0.381$ \\ 
$7$ & $2.575$ & $2$ & $1$ & $0.350$ \\ 
$8$ & $2.943$ & $2$ & $1$ & $0.324$ \\ 
$9$ & $3.311$ & $3$ & $2$ & $0.382$ \\ 
$10$ & $3.679$ & $3$ & $2$ & $0.366$ \\ 
$11$ & $4.047$ & $4$ & $3$ & $0.390$ \\ 
$12$ & $4.415$ & $4$ & $3$ & $0.380$ \\ 
$13$ & $4.782$ & $4$ & $3$ & $0.370$ \\ 
$14$ & $5.150$ & $5$ & $4$ & $0.385$ \\ 
$15$ & $5.518$ & $5$ & $4$ & $0.378$ \\ 
$16$ & $5.886$ & $5$ & $4$ & $0.371$ \\ 
$17$ & $6.254$ & $6$ & $5$ & $0.382$ \\ 
$18$ & $6.622$ & $6$ & $5$ & $0.377$ \\ 
$19$ & $6.990$ & $6$ & $5$ & $0.372$ \\ 
$20$ & $7.358$ & $7$ & $6$ & $0.379$ \\ 
$50$ & $18.390$ & $18$ & $17$ & $0.373$ \\ 
$100$ & $36.790$ & $36$ & $35$ & $0.371$ \\ 
$300$ & $110.400$ & $110$ & $109$ & $0.369$ \\ 
\hline \\[-1.8ex] 
\end{tabular} 

\end{table} 

\footnotetext{See appendix \ref{Appendix B} for R code to obtain values.}

\section{Discussion}
In this paper, we solely focused on the consumers' optimal policy without considerations on the effect they may have on the firm (store)-side. It is reasonable to expect that consumers' behavior will have an impact on the actions of firms, so the next question is; given that consumers use the optimal policy we derived, will there be any changes in the firms' price setting behavior, compared to say the case in \cite{fishman1995durability}? In our analysis we maintained the assumption that stores' prices reflected their productivity state (high cost stores charge higher prices.) However, in \cite{reinganum1979simple} and \cite{fishman1995durability} this was a result of firms behaving optimally given that they were able to deduce the consumers' reservation price $p_r$. Since our search policy does not use a reservation price, there is additional work to be done to see how firms' will behave under the search policy derived here.\footnote{To this end, one may consider using the `adversarial bandit' variant of the MAB introduced by \cite{auer2002nonstochastic}.}

We have also taken quite a different approach to deriving the consumer's optimal search rule. Although upon closer inspection both the consumer's search rule in \cite{fishman1995durability} and our optimal stopping rule work similarly in some aspects.

In \cite{fishman1995durability}, the consumer's search rule is defined by a reservation price $p_r$ where the rule is 
\begin{equation}
    \textit{`If $p_j > p_r$, continue search, otherwise settle for $p_j$.'}
    \label{fishman rule}
\end{equation} 

compared to our optimal stopping rule for a given $M$ number of stores, 
\begin{equation}
    \textit{`Pass through $m^*-1$ stores, after which pick the first store that yields the highest value so far.'}
    \label{secretary policy}
\end{equation}

They are similar in that both policies are characterized by a cut-off; the reservation price $p_r$ in \cite{fishman1995durability} and $m^*$ in ours. The basis of Fishman's search rule is to induce an indifference condition with the reservation price $p_r$ given a persistence probability $\beta$ and search cost $c$. We have seen that changes in $\beta$ indeed influences the consumer's choice, where consumers choose lower reservation prices the more persistent the stores' states are. This is also reflected in the policy we derived. The consumer picking a certain store for its value $V_j$ is also a result of the the transition/persistence probabilities as the value takes the future expectation into account.

To illustrate, suppose the reward the consumer receives $\pi\Big[x_j(t)\Big]$ is indeed high. If this store is likely to continue its current state into the future, this will be reflected in the value $V$ as defined in (\ref{dynamic store values}). Conversely, if the store's state is highly volatile, this will also be reflected in $V$, affecting its desirability to the consumer. 

In the end, we can see that although the policies were formulated very differently, they do share similarities in what they do, in that they both try to make the best decision based on current and expected future payoffs.

\clearpage

\bibliographystyle{apacite}
\bibliography{ref}

\clearpage
\renewcommand\thesubsection{\Alph{subsection}}
\section*{Appendices}
\begin{appendices}
	\titleformat*{\section}{\fontsize{11}{20}\bfseries}
\subsection{Value of the Optimal Stopping Policy}\label{Appendix A}
We can write the optimal stopping policy's value $Y_0 = Y_{m^*-1}$ in terms of the cutoff $m^*$,
\begin{equation}
    \begin{split}
        Y_0 &= \Big(\frac{m^*-1}{m^*}\Big) Y_{m^*} + \Big(\frac{1}{m^*}\Big)U_{m^*} \\[0.2cm]
        &= \Big(\frac{m^*-1}{m^*}\Big) \Big(\frac{m^*}{M}\Big)\sum_{k=m^*}^{M-1}\frac{1}{k} + \frac{1}{m^*}\Big(\frac{m^*}{M}\Big) \\[0.2cm]
        &= \frac{m^*-1}{M}\Big(\frac{1}{m^*}+\frac{1}{m^*+1}+\ldots+\frac{1}{M}\Big)+\frac{1}{M} \\[0.2cm]
        &= \frac{m^*-1}{M}\Big(\frac{1}{m^*-1}+\frac{1}{m^*}+\ldots+\frac{1}{M-1}\Big)\\[0.2cm]
        &= \frac{m^*-1}{M}\sum_{k=m^*-1}^{M-1}\frac{1}{k}.
    \end{split}
    \label{value of optimal policy in m*}
\end{equation}

where the second line in (\ref{value of optimal policy in m*}) uses the fact mentioned above about $m^*$ being the first store to satisfy the condition $m^*/M \leq m^*/M \cdot \sum_{k=m^*}^{M-1}1/k$.

\subsection{R Code for Deriving Policy Values}\label{Appendix B}
Visit the following url for the R code. 

\begin{verbatim}
    https://github.com/ctymarco/MABandSecretaries/blob/b1345a7d2c051129d98bcf82c0 
    81be1434756bc1/values.R
\end{verbatim}

\end{appendices}

\end{document}